\documentclass[aps,prl,twocolumn,superscriptaddress,longbibliography]{revtex4-1}

\usepackage{graphicx}
\usepackage{amsmath}
\usepackage{ulem}
\usepackage{color}
\usepackage[colorlinks=true,linkcolor=blue,urlcolor=blue,citecolor=blue]{hyperref}
\usepackage{braket}
\usepackage{amssymb}
\usepackage{commath}

\begin{document}


\title{Coupling Identical 1D Many-Body Localized Systems}

\author{Pranjal Bordia}
\affiliation{Fakult\"at f\"ur Physik, Ludwig-Maximillians-Universit\"at M\"unchen, Schellingstr. 4, 80799 Munich, Germany}
\affiliation{Max-Planck-Institut f\"ur Quantenoptik, Hans-Kopfermann-Str. 1, 85748 Garching, Germany}

\author{Henrik P. L\"uschen}
\affiliation{Fakult\"at f\"ur Physik, Ludwig-Maximillians-Universit\"at M\"unchen, Schellingstr. 4, 80799 Munich, Germany}
\affiliation{Max-Planck-Institut f\"ur Quantenoptik, Hans-Kopfermann-Str. 1, 85748 Garching, Germany}

\author{Sean S. Hodgman}
\affiliation{Fakult\"at f\"ur Physik, Ludwig-Maximillians-Universit\"at M\"unchen, Schellingstr. 4, 80799 Munich, Germany}
\affiliation{Max-Planck-Institut f\"ur Quantenoptik, Hans-Kopfermann-Str. 1, 85748 Garching, Germany}
\affiliation{Research School of Physics and Engineering, Australian National University, Canberra ACT 0200, Australia}

\author{Michael Schreiber}
\affiliation{Fakult\"at f\"ur Physik, Ludwig-Maximillians-Universit\"at M\"unchen, Schellingstr. 4, 80799 Munich, Germany}
\affiliation{Max-Planck-Institut f\"ur Quantenoptik, Hans-Kopfermann-Str. 1, 85748 Garching, Germany}

\author{Immanuel Bloch}
\affiliation{Fakult\"at f\"ur Physik, Ludwig-Maximillians-Universit\"at M\"unchen, Schellingstr. 4, 80799 Munich, Germany}
\affiliation{Max-Planck-Institut f\"ur Quantenoptik, Hans-Kopfermann-Str. 1, 85748 Garching, Germany}

\author{Ulrich Schneider}
\affiliation{Fakult\"at f\"ur Physik, Ludwig-Maximillians-Universit\"at M\"unchen, Schellingstr. 4, 80799 Munich, Germany}
\affiliation{Max-Planck-Institut f\"ur Quantenoptik, Hans-Kopfermann-Str. 1, 85748 Garching, Germany}
\affiliation{Cavendish Laboratory, University of Cambridge, J. J. Thomson Avenue, Cambridge CB3 0HE, United Kingdom}

\date{\today}

\begin{abstract}
We experimentally study the effects of coupling one-dimensional Many-Body Localized (MBL) systems with identical disorder. Using a gas of ultracold fermions in an optical lattice, we artifically prepare an initial charge density wave in an array of 1D tubes with quasi-random onsite disorder and monitor the subsequent dynamics over several thousand tunneling times. We find a strikingly different behavior between MBL and Anderson Localization. While the non-interacting Anderson case remains localized, in the interacting case any coupling between the tubes leads to a delocalization of the entire system.
\end{abstract}


\maketitle


\textit{Introduction.---}Many-Body Localization (MBL) marks a new paradigm in condensed matter and statistical physics. It describes an insulating phase in which a disordered, interacting many-body quantum system fails to act as its own heat bath~\cite{Deutsch91,Sred94,Rigol08,RMPPolkovnikov11,Nandkishore15}. In isolation, these systems will never achieve local thermal equilibrium and conventional statistical physics approaches break down. Unlike other insulating phases, MBL is not limited to ground states but can even occur in all exited states of a disordered many-body system~\cite{Basko06,Polyakov05,Oganesyan05,Imbrie14,Shlyapnikov15}. A dynamical phase transition separates the MBL phase from conventional ergodic phases~\cite{Pal10,Vosk13}, in which the isolated system thermalizes. In these ergodic phases, any initial quantum information becomes rapidly diluted in the exponentially large Hilbert space, leading to decoherence. In contrast, in the localized insulating phase quantum information can persist locally for an infinite amount of time~\cite{Nandkishore15}. This could potentially render quantum information devices less susceptible to noise and disorder. For many decades it remained unclear whether such a localized phase could persist in a many-body system beyond the non-interacting limit of an Anderson insulator~\cite{Anderson58}. Today, both theory and experiment have shown evidence for the existence of an MBL phase in interacting 1D systems~\cite{Basko06,Schreiber15,Polyakov05,Nandkishore15,Altman15,Monroe15}. Nonetheless, many fundamental questions regarding this phase and the associated phase transition as well as its extension to higher dimensions ~\cite{Yao15} remain open, making it a highly active topic of current theoretical and experimental research.

One crucial requirement for the existence of a MBL phase is that no coupling to any external heat bath or bath-like structure exists. Any such coupling will eventually thermalize the system and ultimately destroy the non-thermalizing MBL phase~\cite{Mott69,Nandkishore14,Johri15,Nandkishore15,Huse15,Garrahan15}. Since any experimental system will inevitably be coupled---albeit potentially very weakly---to an environment, it is of critical importance to quantitatively understand the effect of such a coupling. Furthermore, studying the effects of weak couplings can help to experimentally identify an MBL phase and distinguish it from glasses or non-interacting Anderson localized phases.


\begin{figure}
\centering
\includegraphics[width=84mm]{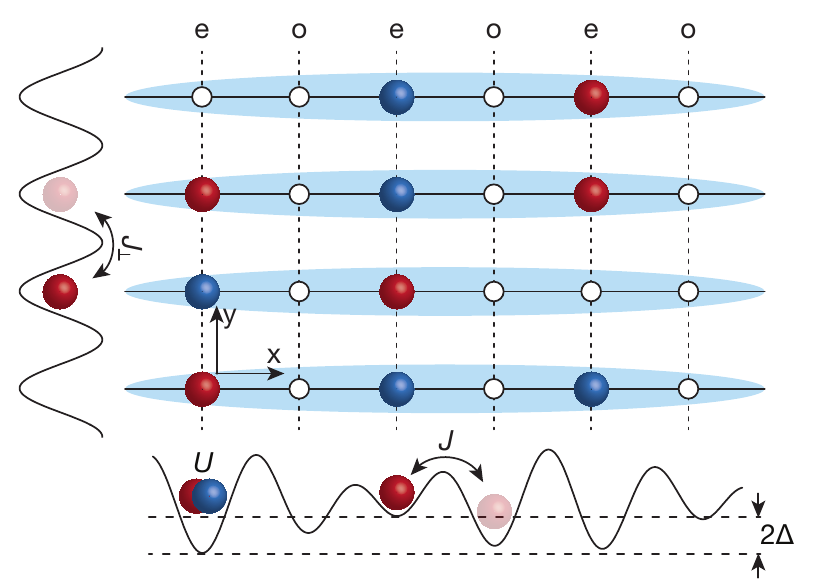}
\caption{\small{Coupling identical MBL systems: A charge density wave (CDW) with atoms only occupying even sites ($e$) is prepared in each of the identically disordered 1D tubes along the longitudinal ($x$) direction, with hopping $J$, on-site interaction energy $U$ and disorder strength $\Delta$. Red and blue spheres indicate a typical distribution of $\left|\uparrow\right\rangle$ and $\left|\downarrow\right\rangle$ atoms. We monitor the time evolution of such a state for different inter-tube coupling strengths $J_{\perp}$, that is different hopping amplitudes along the transverse ($y$) direction.}}
\label{Schematic_fig}
\end{figure}
Ultracold quantum gases in optical lattices form an ideal system to investigate these questions, as they are almost perfectly isolated from the external world and are highly controllable. Earlier experiments have investigated the interplay between disorder and interactions in the ground state of an isolated systems of ultracold bosons in optical lattices~\cite{Deissler10,Gadway11,Errico14} and studied the influence of disorder on transport properties for lattice fermions~\cite{Kondov15}. Recently, we were able to show that for a wide range of energy-densities and interactions, an MBL phase exists in 1D Hubbard type chains with quasi-random disorder~\cite{Schreiber15}.

In this work, we experimentally study the effects of coupling identically disordered 1D MBL systems to each other. In particular, the disordered Hamiltonian is identical for all 1D tubes, but the initial configuration of atoms differs between the tubes (Fig.~\ref{Schematic_fig}). We find that the coupled systems can collectively serve as a bath for each other, i.e.\ coupling localized systems can result in delocalizing all of them.


\textit{Experiment.---}Our experiments start with a two component Fermi gas of $^{40}$K atoms in an equal mixture of the two lowest hyperfine states $\left|F,m_F\right\rangle = \left|\frac{9}{2},-\frac{9}{2}\right\rangle \equiv \left|\downarrow\right\rangle$ and $\left|\frac{9}{2},-\frac{7}{2}\right\rangle \equiv \left|\uparrow\right\rangle$ with a total atom number of about $110$-$150\times10^{3}$ atoms. In the initial dipole trap, the atoms are at a temperature of $0.19(2)$\,$T_F$, where $T_F$ is the Fermi temperature. We load the Fermi gas into the lowest band of a deep, three-dimensional simple cubic optical lattice, where tunneling can be neglected. Along the longitudinal ($x$) direction, we then add a second (short) lattice (wavelength $\lambda_s$=532\,nm) to the initial (long) lattice ($\lambda_l$=1064\,nm). By controlling the phase of the short lattice during loading, we prepare a period-two 'Charge-Density-Wave' (CDW), where only even sites are occupied in the 20\,$E_r$ deep short lattice. Here, the recoil energy is denoted by $E_{r} = h^2/2m\lambda^2$, where $h$ is Planck's constant, $\lambda$ is the respective lattice wavelength and $m$ is the atomic mass. The orthogonal lattices along $y$ and $z$ with a wavelength of $\lambda_{\perp}$=738\,nm are initially ramped up to 45\,$E_{r}$, creating an array of (almost) isolated 1D tubes. During lattice loading the interactions are kept strongly repulsive at a scattering length of $a$ = 142\,a$_\text{0}$, where a$_\text{0}$ is the Bohr radius, by employing a Feshbach resonance centered at 202.1\,G~\cite{Regal03}. This results in a doublon fraction, that is the fraction of atoms on doubly occupied lattice sites, of $\leq 10\%$. 

\begin{figure}[tb]
\centering
\includegraphics[width=84mm]{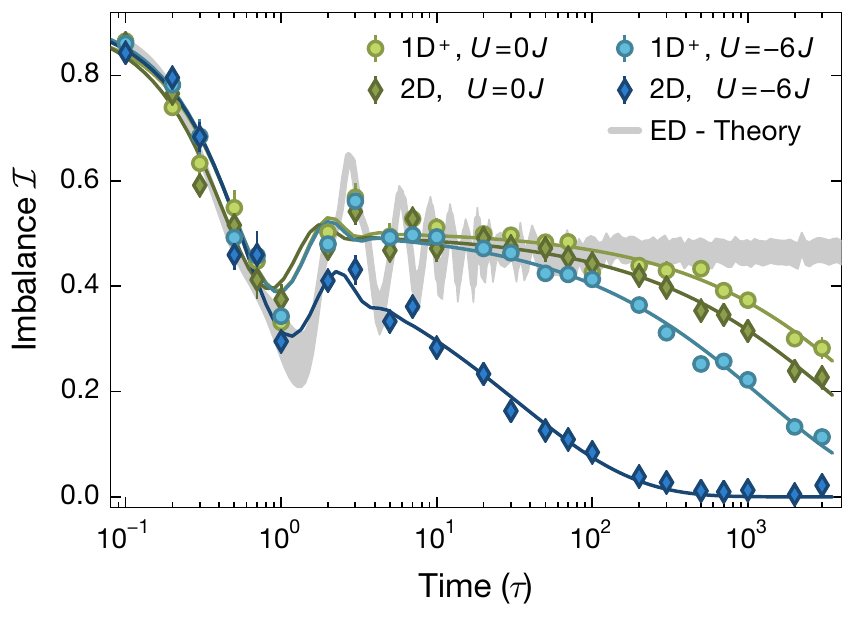}
\caption{\small{Time evolution of a charge-density wave (CDW): An initially prepared 1D CDW evolves in a disordered system with disorder strength $\Delta=5\,J$ along one axis. We measure the imbalance $\mathcal{I}$ after a given evolution time in 1D$^+$ (circles), where $J_\perp \lesssim 10^{-3} J$ and in 2D (diamonds), where $J_\perp = J$. Each data point is the average of 6 disorder phase realizations, with error bars showing the standard error of the mean. Solid lines are fits~\cite{SOMs} from which we extract the imbalance lifetimes. Shown in gray is an exact diagonalization (ED) calculation for $J_{\perp}=0$, $\Delta=5$\,$J$ and $U=0$~\cite{SOMs}.}}
\label{time_trace_fig}
\end{figure}

After the preparation of the CDW in the deep lattices, the desired interactions for the ensuing evolution are set. Additionally, an incommensurate lattice of wavelength $\lambda_{d}$=738\,nm is superimposed along the $x$ direction to create quasi-random onsite disorder along the longitudinal direction. The system size is approximately 200 sites in the longitudinal and 120 sites in the transverse direction, and the central longitudinal tubes contains about 90 atoms~\cite{SOMs}. After this preparation, the long lattice is quickly ramped to zero, the transverse $y$ lattice is ramped to its final value, which controls the transverse coupling $J_\perp$, and the short lattice is reduced to $8\,E_r$. This last ramp enables tunneling along the tube and thereby initiates the dynamics. After a variable evolution time, we extract the imbalance between atoms on even and odd sites $\mathcal{I}=(N_{e}-N_{o})/(N_{e}+N_{o})$~\cite{Schreiber15}.  Here $N_{e}$ and $N_{o}$ denote the population of even and odd sites respectively along the longitudinal direction and are extracted by mapping them to different bands of the superlattice~\cite{Trotzky12}.

The imbalance provides a measure of ergodicity breaking: It quickly decays to zero under any ergodic dynamics and any non-zero imbalance persisting at long times signifies a memory of the initial state and directly indicates localization~\cite{Schreiber15,Rigol15,Garrahan15}.

\textit{Model.---}Our system consists of an array of 1D tubes with identical quasi-random on-site disorder along the longitudinal direction.  Each tube can be described by the Aubry-Andr{\'e} Model~\cite{Aubry80} with interactions~\cite{Iyer13}, as depicted in Fig.\ \ref{Schematic_fig}. A finite hopping amplitude $J_{\perp}$ along the transverse ($y$) direction introduces a coupling between adjacent tubes. The Hamiltonian of the system is given by

\begin{equation}
\begin{split}
	\hat{H} = &-J\sum_{i,j,\sigma}(\hat{c}^{\dagger}_{i+1,j,\sigma}\hat{c}_{i,j,\sigma} + \text{h.c.}) \\&  -J_{\perp}\sum_{i,j,\sigma}(\hat{c}^{\dagger}_{i,j+1,\sigma}\hat{c}_{i,j,\sigma} + \text{h.c.}) \\& + \Delta\sum_{i,j,\sigma} \cos (2\pi\beta i+\phi)\hat{n}_{i,j,\sigma} +U\sum_{i,j} \hat{n}_{i,j,\uparrow}\hat{n}_{i,j,\downarrow},
	\label{AA_hamiltonian}
\end{split}
\end{equation}

\noindent
where $J\approx h\times 500$\,Hz is the tunneling matrix element between neighboring sites along a tube and $J_{\perp}$ denotes the transverse ($y$) hopping between the tubes. The creation (annihilation) operator for a fermion in spin state $\sigma\in\{\uparrow,\downarrow\}$ on site $i$ in tube $j$ is $\hat{c}_{i,j,\sigma}^{\dagger}$($\hat{c}_{i,j,\sigma}$) and the local number operator is given by $\hat{n}_{i,j,\sigma} = \hat{c}_{i,j,\sigma}^{\dagger}\hat{c}_{i,j,\sigma}$. The quasi-random on-site disorder is characterized by the disorder amplitude $\Delta$, the incommensurable wavelength ratio $\beta = \lambda_{s} / \lambda_{d}$ and the relative phase $\phi$. Finally, the on-site interaction energy is given by $U$.

In the limit $J_{\perp}\rightarrow 0$, the system decouples into many one-dimensional tubes, which show many-body localization~\cite{Schreiber15}. For our experiment, the accessible limits of almost zero inter-tube coupling $J_{\perp} \lesssim 10^{-3}$\,$J$ and equal coupling $J_{\perp} = J$ are termed the 1D$^+$ and 2D cases, respectively. Note that the experimentally achievable limit of the 1D$^+$ case is \textsl{not} the same as the ideal theoretical 1D case. A small but non-zero coupling remains for any finite transverse lattice depth and can always affect the dynamics at very long time scales.

\textit{Results.---}We monitor the time evolution of the imbalance at disorder amplitude $\Delta=5$\,$J$ for various interaction strengths $U$ in both the 1D$^+$ and 2D cases. This disorder strength is deep in the MBL regime for isolated 1D tubes.  Fig.\ \ref{time_trace_fig} shows exemplary time traces at $U=0$ and $U=-6$\,$J$, with all times given in units of the longitudinal tunneling time $\tau = \hslash/J$. We start with an out-of-equilibrium density wave with an initial imbalance of $\mathcal{I}(t=0)=0.91\pm0.03$ and observe a fast initial decrease of the imbalance up to approximately one tunneling time. This decrease is similar in all cases and corresponds to an initial relaxation in the longitudinal direction. 
\begin{figure}
\centering
\includegraphics[width=84mm]{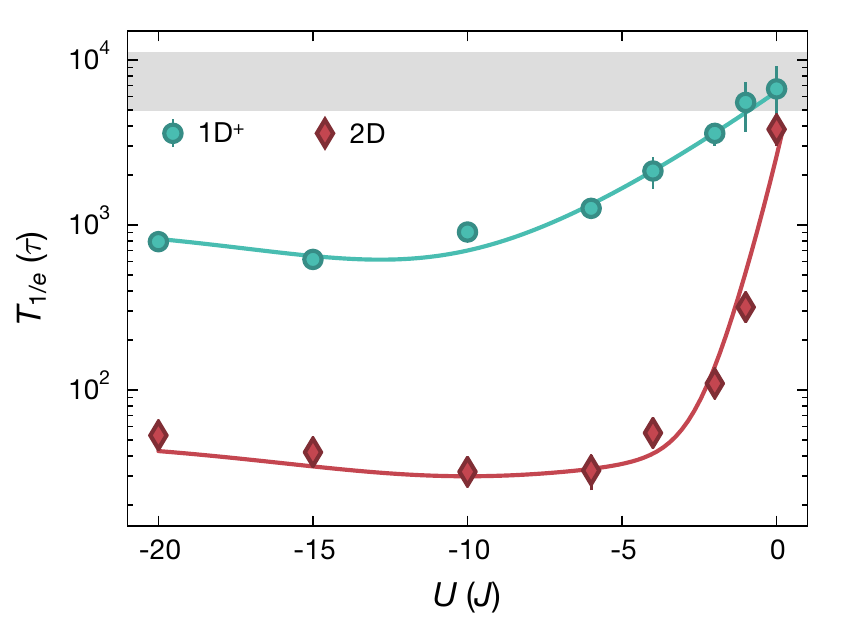}
\caption{\small{Imbalance lifetimes versus interactions ($U$): Imbalance lifetimes at $\Delta=5$\,$J$ for 1D$^+$ and 2D. We note that the 1D$^+$ case differs crucially from the ideal isolated 1D case. The lifetimes were extracted from fits to time traces such as in Fig.\ \ref{time_trace_fig}. Error bars denote fit uncertainty ~\cite{SOMs}. The gray shaded area indicates the range of measured atom number lifetimes, while the lines are guides to the eye.}}
\label{lifetime_U_fig}
\end{figure}

\begin{figure}
\centering
\includegraphics[width=84mm]{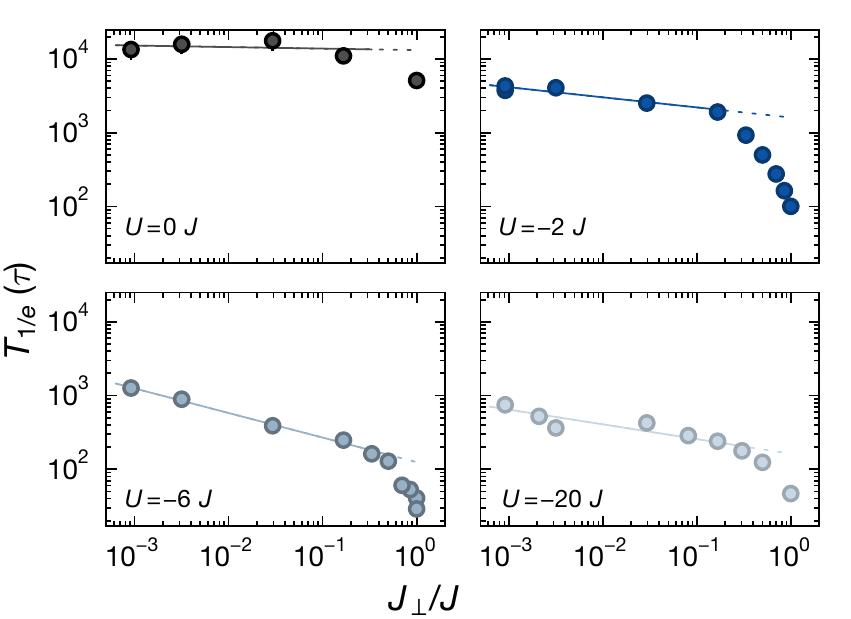}
\caption{\small{Imbalance lifetimes in the 1D-2D crossover: The coupling strength $J_{\perp}$ between adjacent tubes is varied continuously for $\Delta=5\,J$ at four different interactions $U$. Here $J_{\perp}/J \lesssim 10^{-3}$ and $J_{\perp}/J=10^{0}$ correspond to the 1D$^+$ and the 2D cases respectively. The lack of saturation as $J_{\perp} \rightarrow 0$ indicates that the residual inter-tube coupling still limits the imbalance lifetime in the 1D$^+$ case. Solid lines denote power-law fits $\propto J_{\perp}^k$ for small $J_{\perp}$, with fitted exponents $k(U/J)$ of $k(0)=0.00(4)$, $k(-2)=-0.09(2)$, $k(-6)=-0.30(1)$ and $k(-20)=-0.16(3)$. We note that in principle the tunneling along the $z$ direction becomes sizable for the smallest $J_{\perp}$, as $J_{\perp}^z/J \sim 10^{-3}$.}}
\label{crossover_summary_fig}
\end{figure}

As shown in Fig.\ \ref{time_trace_fig}, in both non-interacting cases, the initial decrease is followed by highly damped oscillations around a plateau at finite imbalance, closely matching the expected steady state for this Anderson localized system~\cite{Schreiber15,SOMs}. At very long times ($\gg 100\,\tau$), the curves start to deviate from this plateau and exhibit a slow decay. The corresponding lifetime is extracted by fitting the imbalance traces to a damped sinusoid, which models the initial fast relaxation, multiplied by a stretched exponential to capture the slow decay~\cite{SOMs}. This stretched exponential is of the form $e^{-(\Gamma t)^\beta}$, where $t$ is the evolution time, $\Gamma$ is the decay rate and $\beta$ is the stretching exponent. This model fits all time traces consistently better than a simple exponential decay and has also been observed in glasses, disordered materials and polymers \cite{Orbach84, Reisfeld84, Philipps96}. We define our imbalance lifetime $T_{1/e} = 1/\Gamma$.

In the absence of interactions, the observed long time dynamics is dominated by classical noise, photon scattering from lattice beams and other technical imperfections. These processes couple the system to the environment and over time delocalize it~\cite{Gurvitz2000,Nowak2012}. In addition, these experimental imperfections also give rise to an atom number decay, which limits the lifetime of atoms in the lattice to  $0.5-1.1 \times 10^{4}$\,$\tau$~\cite{SOMs}. In the absence of interactions, the observed dynamics in 1D$^+$ and 2D are almost the same, since the disorder potential is identical in all tubes~\cite{SOMs}. Therefore the 2D system is separable, and the longitudinal and transverse directions are decoupled. Note that we do not expect this to hold if the disorder were different in different tubes~\cite{Abrahams79}.  
Adding interactions, however, breaks the separability of the system and the transverse dynamics can now affect the imbalance along the longitudinal direction. Since there is no disorder along the transverse direction, particles are free to move along this direction. In the interacting case, this couples the originally localized tubes such that they collectively act as a bath for each other and thereby delocalize the entire system. As a consequence, the interacting 2D trace in Fig.~\ref{time_trace_fig} displays no plateau but instead shows a fast decay. To understand this behavior further, we map out the imbalance lifetimes in the 1D$^+$ and 2D cases for various interaction strengths (Fig.~\ref{lifetime_U_fig}). We find that in 2D, even small interactions are sufficient to dramatically reduce the imbalance lifetime to less than $100\, \tau$.

Although MBL is expected to be stable in the isolated 1D case, the 1D$^+$ data in Fig.~\ref{lifetime_U_fig}  shows qualitatively similar behavior to the 2D case, but with a much weaker decrease of imbalance lifetimes with increasing interactions. This suggests that the small but non-zero inter-tube coupling also limits the lifetime in the 1D$^+$ case. The gray shaded region in Fig.~\ref{lifetime_U_fig} marks the range of observed atom number lifetimes, which approximately coincides with the non-interacting imbalance lifetimes, indicating the relevance of technical imperfections on this timescale.


\textit{1D-2D Crossover.---}In order to directly test the effect of residual couplings, we vary the strength of the inter-tube coupling between the 2D and the 1D$^+$ limits for four different interactions (Fig.\ \ref{crossover_summary_fig}). We observe increasing lifetimes for decreasing coupling strengths $J_{\perp}$ in all interacting cases. For small but finite coupling strengths, we observe a linear trend on a log-log scale (Fig.\ \ref{crossover_summary_fig}), suggesting a power law dependence. For strong inter-tube coupling ($J_{\perp}/J \gtrsim 0.1$), there is a crossover to a faster decay. The fitted exponents are surprisingly small ($\abs{k}<1$) and depend non-monotonically on the interactions. In the non-interacting limit $U\rightarrow 0$, the lifetimes are on the order of the atom number lifetime and become independent of the transverse coupling due to the separability of the problem. This highlights the striking difference between MBL and Anderson localization.


Extrapolating towards the limit of the true 1D case ($J_{\perp} \rightarrow 0$), we would expect the lifetimes of the interacting system to saturate once the inter-tube coupling is no longer the dominant decay mechanism. However, in the experimentally accessible regime we cannot observe any signs of saturation, strongly suggesting that the interacting lifetimes are still limited by the non-zero residual transverse coupling to the neighboring tubes, even in the 1D$^+$ case.


\begin{figure}
\centering
\includegraphics[width=84mm]{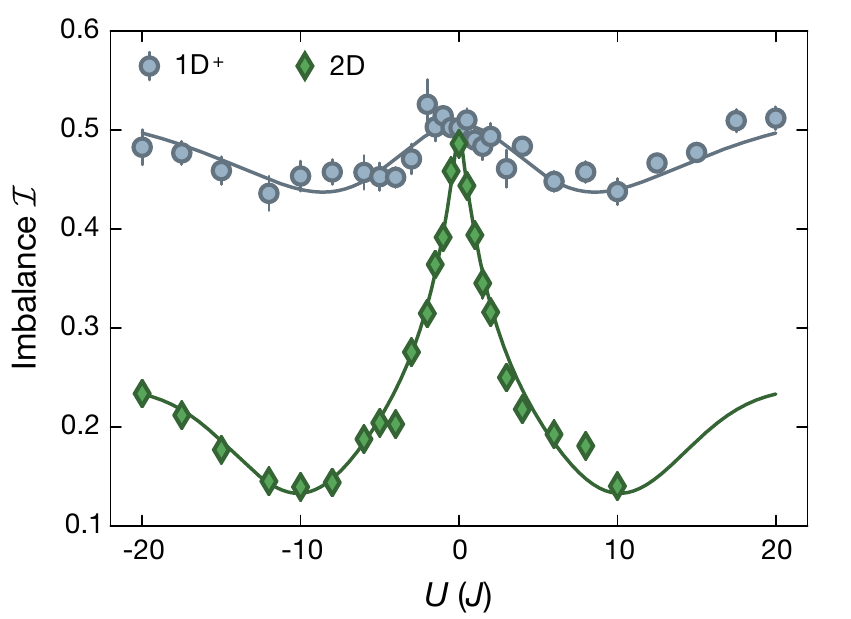}
\caption{\small{Imbalance vs.\ interactions ($U$) at fixed times: Data taken at $\Delta=5\, J$ averaged over three different times ($38 \, \tau-41 \, \tau$) and 4 disorder phases $\phi$. In the 1D$^+$ case (circles), the imbalance corresponds to the stationary value~\cite{Schreiber15}, whereas in the 2D case (diamonds) it is indicative of the decay lifetime. Solid lines are guides to the eye.  The 2D case is limited to $U/J \leq 10$ due to details of the used Feshbach resonance~\cite{Regal03}.}}
\label{const_times_fig}
\end{figure}

\textit{Constant evolution time.---}The interaction dependence is also visible in the imbalance value measured after a fixed evolution time of around $40 \, \tau$, as shown in Fig.~\ref{const_times_fig}. We observe a substantial decrease of the 2D imbalance with small interactions, which is consistent with the sharply decreasing lifetimes. Additionally, Fig.~\ref{const_times_fig} shows that the dynamics are symmetric around $U=0$, which is expected due to a dynamical symmetry of the Fermi-Hubbard Model~\cite{Schneider12}. Interestingly, we observe an increasing imbalance in the 2D data for strong interactions. We checked that this increase is not due to doublons by removing any residual doublons with a pulse of near-resonant light~\cite{Ronzheimer13} prior to the evolution. This increase in lifetime might be due to the reduced phase space for scattering in the hard-core limit ($U \gg J$). A recent theoretical study using a cluster expansion method on a smaller system (8 $\times$ 8 sites) observed a similar trend~\cite{Reichl15}.

At $40$\,$\tau$, the 1D$^+$ case is dominated by the 1D plateau value and shows the characteristic `W' shape of the steady state imbalance of the MBL system~\cite{Schreiber15}, with little influence from the (much longer) lifetimes. As shown in the supplementary material~\cite{SOMs}, we additionally find that the imbalance lifetime in all cases increases strongly for larger disorder.

\footnotetext[3]{We also measured full time traces for $U=+10 \, J$ and observe that they are the same as for $U=-10 \, J$. See Supplementary Material for the full time trace.} 

\textit{Conclusion.---}We have studied the stability of many-body localized 1D systems under an inter-tube coupling. We found that even weak couplings have a delocalizing effect on the MBL phase, while leaving the non-interacting Anderson limit unchanged. This highlights the differences between these two regimes and shows the principal fragility of MBL with respect to coupling to any external heat bath~\cite{Mott69,Nandkishore14,Johri15,Nandkishore15,Huse15}. Furthermore, we have not observed any saturation in imbalance lifetimes even for the smallest couplings we could attain, indicating that this inter-tube coupling is the dominant decay mechanism in this MBL experiment. Nonetheless, for strong disorders the achieved lifetimes already exceed the lifetimes of typical many-body states, such as e.g.\ superfluid states in optical lattices. This demonstrates the stability of MBL with respect to other experimental imperfections and increases the prospects of realizing localization protected order~\cite{Huse13,Bauer13,Chandran14} and applications in quantum-information~\cite{Nandkishore15}.

An important next step will be to extend this study to the `true' 2D case with disorder along both directions. In addition, future experiments should also be able to address the question of the stability of MBL under external influences, such as photon scattering~\cite{Gurvitz2000,Nowak12} and time-dependent modulations~\cite{Gopalakrishnan15,Agarwal15}. Finally, it would also be interesting to search for MBL in bosonic systems~\cite{Aleiner10} and using other observables such as the growth of entanglement entropy and response to generalized interferometric probes~\cite{Serbyn14} when a bath is added to an otherwise perfectly MBL system~\cite{Garrahan15}.

\begin{acknowledgments}
We acknowledge useful discussions with E. Altman, E. Demler,  M. Fischer, D. Huse, M. Knapp, R. Nandkishore and A. Pal. We acknowledge financial support by the European Commision (UQUAM, AQuS) and the Nanosystems Initiative Munich (NIM).
\end{acknowledgments}

\footnotetext[1]{See Supplementary Material for details}
\bibliography{MBL_higher_dim}

\cleardoublepage

\appendix

\setcounter{figure}{0}
\setcounter{equation}{0}

\renewcommand{\thepage}{S\arabic{page}} 
\renewcommand{\thesection}{S\arabic{section}}  
\renewcommand{\thetable}{S\arabic{table}}  
\renewcommand{\thefigure}{S\arabic{figure}} 
\section{Supplementary Material}

\begin{figure}
\centering
\includegraphics[width=84mm]{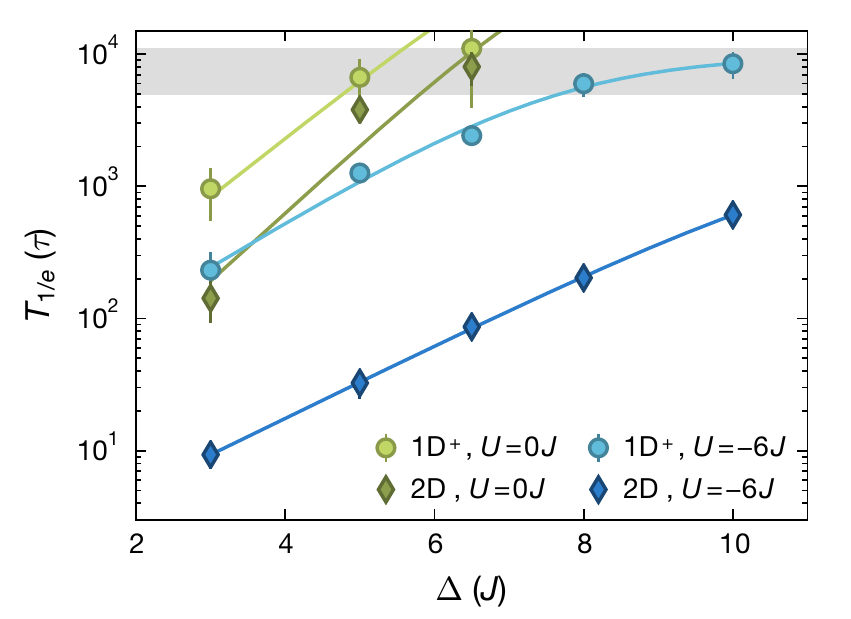}
\caption{\small{Imbalance lifetimes as a function of disorder strength: The lifetimes were extracted by fits to individual time traces taken up to $3000\,\tau$, as in the main text. Error bars indicate uncertainty in the fitting procedure. The gray shaded area indicates the range of typical atom number lifetimes. The lifetimes for the non-interacting cases at larger disorder strengths exceed the atom number lifetimes and hence cannot be reliably extracted.} Solid lines show the results of saturated exponential fits as described in the text.}
\label{lifetime_delta_fig}
\end{figure}

\textit{Imbalance Lifetimes vs Disorder Strength.---}Fig.~\ref{lifetime_delta_fig} shows the dependence of imbalance lifetimes on the disorder strength $\Delta$ in the 1D$^+$ and the 2D case for $U=-6 \, J$. In both cases, we observe exponentially increasing lifetimes with larger $\Delta$. We attribute this to a reduced effectiveness of any bath-like perturbation for large disorder strengths. At long lifetimes this behavior saturates, as the imbalance lifetimes approach the same order of magnitude as the atom loss lifetimes. Since all time traces in the paper are measured up to $3000 \, \tau$, lifetimes exceeding $5000 \, \tau$ may be affected by systematic errors due to the fitting procedure becoming unreliable.

To capture the behavior with $\Delta$ we fit the data with the following empirical function:
 
\begin{equation} \label{fit_func_Imb}
T_{1/e} = \frac{1}{\Gamma_{b}+ae^{-b\Delta}},
\end{equation}

where $\Gamma_{b}$ represents the background imbalance loss due to noise, while $a$ and $b$ are free parameters. These fits are represented as solid lines in Fig.\ \ref{lifetime_delta_fig}. The fit routine is performed in logarithmic units.
\\

\textit{Non-Interacting Lifetime Difference between 1D$^+$ and $2D$.---}The difference between the non-interacting 1D$^+$ and the 2D data is most likely due to technical imperfections. Firstly, the interactions $U$ may not be exactly zero for all times as it relies on the stability and the absolute calibration of the magnetic field. Secondly, the lattice beams have gaussian envelopes. This implies that both longitudinal hopping and disorder depend slightly on the transverse position. Thirdly, the relative phase of the disorder might be slightly different between neighboring tubes due to slight misalignment between the disorder beam and the main lattice. These imperfections break the separability of the problem. 

\begin{figure}
\centering
\includegraphics[width=84mm]{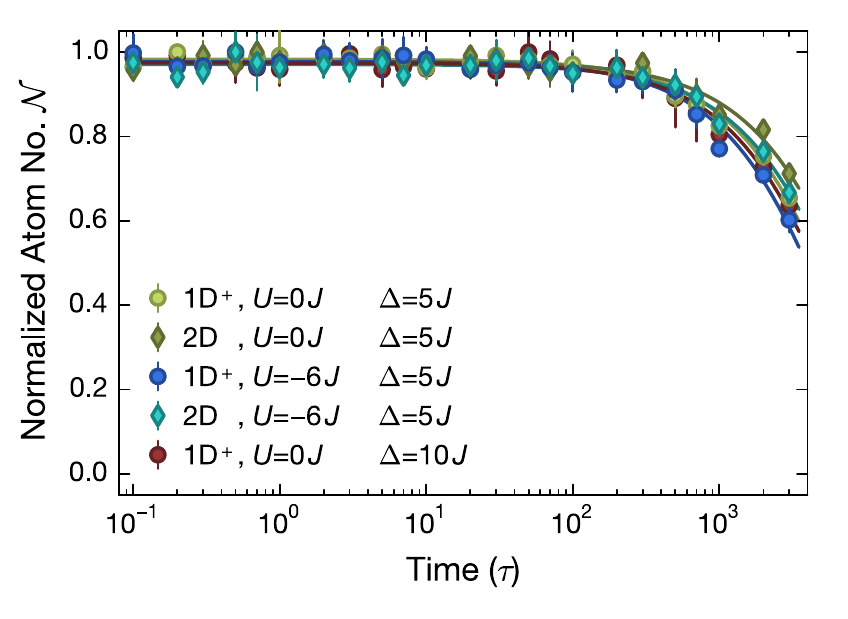}
\caption{\small{Normalized atom loss time traces: Characteristic traces showing the loss of atoms for different couplings $J_{\perp}$, disorders $\Delta$ and interactions $U$. Every data point is averaged over 6 disorder realizations $\phi$. Each trace was individually normalized to its maximum value after averaging. Error bars show the error of the mean. Lines are fit with exponentials of the form $Ae^{-t/T_{a}}$.}}
\label{atom_loss_fig}
\end{figure}

\textit{Atom Number Lifetimes.---}In order to determine the atom number lifetime, we fit an exponential decay to the measured atom number time trace for every imbalance dataset. Five typical curves are shown in Fig.\ \ref{atom_loss_fig}. The lifetimes are similar across the full investigated parameter range and we see no strong trend with either $\Delta$ or $U$. Extracted lifetimes from the exponential fits lie in the range of $0.5-1.1\times 10^4 \, \tau$.

As the experimental lifetime of atoms in the pure dipole trap is over $10^5 \, \tau$, the detected atom loss in the lattice must be dominated by off-resonant photon scattering of lattice photons as well as frequency and amplitude noise in the optical fields. This is consistent with the extracted atom number lifetimes being slightly higher in the 2D cases as compared to the 1D$^+$ cases, as the total optical power in the lattice beams is lower in the 2D cases. The loss channels formed by such processes will most likely involve excitations of atoms to higher bands. We note that, due to the long observed atom number lifetimes, these loss channels are unlikely to be relevant for the fast imbalance decay at the observed timescales in 2D: Previous DMRG simulations in 1D~\cite{Schreiber15} agreed well with the experimental measurements in the 1D$^+$ case until $\approx100\, \tau$. Hence, we do not expect higher bands to play a significant role up to at least this time. Furthermore, the atom number lifetimes in the 2D cases are even slightly longer than in the 1D$^+$ cases. Besides, at a relatively short time of $40\, \tau$, as presented in Fig.~\ref{const_times_fig}, we already find a strong effect of the inter-tube coupling, despite the fact that at these time scales there is no discernible atom loss.
\\

\begin{figure}
\centering
\includegraphics[width=84mm]{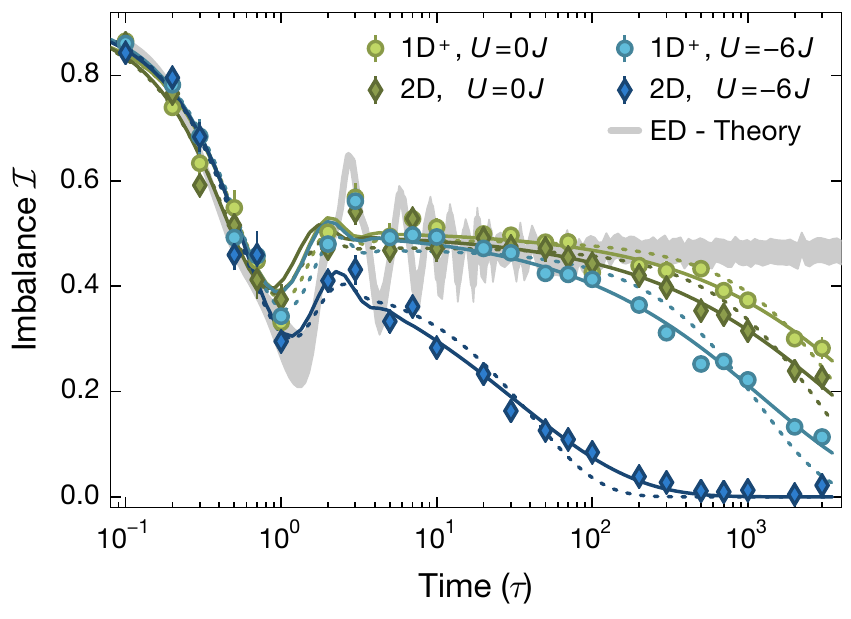}
\caption{\small{We reproduce the time traces from Fig.~\ref{time_trace_fig} in the main paper with both the stretched exponential (solid lines, eq.~\ref{fit_func_Imb}) and a pure exponential fit (dotted lines) demonstrating that the former describes the measured data better. We find this behavior to hold for all the time traces we have measured.}}
\label{time_traces_with_exp}
\end{figure}

\textit{Extracting Imbalance Lifetimes.---}The observed time evolutions of the imbalance $\mathcal{I}$ displays some common characteristic features. The imbalance starts close to unity, followed by a couple of oscillations with a period on the order of $\tau$, which quickly settle to a non-zero plateau value, before decaying to zero in the long-term limit. To quantify this long term decay we fit the sum of a damped sinusoidal and an offset multiplied with an overall stretched exponential decay. This fitting function is given by:

\begin{equation} \label{fit_func_Imb}
\mathcal{I} = (Ae^{-t/t_1}\text{cos}(\nu t) + I_{st})e^{-(\Gamma t)^{\beta}}
\end{equation}

\noindent
where A is the amplitude of the sinusoid with oscillation frequency $\nu \sim 1/\tau$, oscillation damping time $t_1 \sim \tau$, plateau value $I_{st}$, long-term decay rate $\Gamma$ and stretching coefficient $\beta$. From this fit function we obtain the lifetimes $T_{1/e} = 1/\Gamma$, as shown in Figs.~\ref{lifetime_U_fig},~\ref{crossover_summary_fig} and~\ref{lifetime_delta_fig}. This lifetime would diverge in the case of a perfectly localized system without external noise and strictly vanishing bath couplings. The error bars in $T_{1/e}$ throughout the paper are fit uncertainties. We fit a stretched exponential instead of a pure exponential as it captures the measured time traces better. This is illustrated in Fig.~\ref{time_traces_with_exp}. Note that the choice of the stretched or the pure exponential does not qualitatively change any the observed behavior.\\

\noindent


\textit{Exact Diagonalization.---}The ED calculations are adapted from our previous work. We simulate the Anderson localization limit by performing Exact Diagonalization (ED) on a single, non-interacting particle and account for our experimental setup by averaging over fifty phases (equally spaced between $[0,2\pi]$), the dipole trap, the inhomogeneity of the lattice and the atomic cloud shape. Full details of the ED can be found in Ref.~\cite{Schreiber15}.\\

\textit{General Sequence.---}The experiment produces an ultracold gas of fermionic Potassium-40 ($^{40}$K) atoms by sympathetically cooling $^{40}$K with bosonic Rubidium-87 ($^{87}$Rb) in a plugged quadrupole trap followed by an optical dipole trap. Reducing the dipole trap depth lower than a threshold value completely removes $^{87}$Rb due to its higher mass, such that only $^{40}$K remains in the trap. We further evaporate $^{40}$K in an equal mixture of the two lowest hyperfine states of the $^4S_{1/2}$  manifold ($\ket{F,m_F} = \ket{9/2,-9/2} \equiv \ket{\downarrow}$ and $\ket{9/2,-7/2} \equiv \ket{\uparrow}$) to a final temperature of $T/T_F = 0.19(0.02)$, where $T_F$ is the Fermi temperature. Interactions between the two states can be tuned via an s-wave Feshbach resonance centered at 202.1G~\cite{Regal03}. The scattering length $a$ is set to $140 \, \text{a}_{\text{0}}$ during lattice loading to avoid double occupancies. Here, $\text{a}_{\text{0}}$ denotes the Bohr radius. The number of double occupancies are characterized by converting them into Feshbach molecules via crossing the Feshbach resonance and subsequently imaging the remaining single atoms~\cite{Schneider08}. 

After loading into the deep lattices (as described in the main text), we then set the scattering length to control the desired interactions $U$ in the lattice for the following evolution time. At the end of this preparation stage, we ramp down the $x$-short lattice from $20 \, E_r$ to $8 \, E_r$ in $10 \, \mu$s and ramp down the $x$-long lattice to $0\,E_r$in the same time. The disorder lattice is simultaneously ramped up to the desired disorder strength. The orthogonal $y$-lattice is ramped to the final value with an additional time of $90 \, \mu$s. All of these lattice ramp times are short compared to a tunneling time $\tau$. The system is then allowed to evolve for various evolution times. For detection, the short lattice, the long lattice and the orthogonal lattices are ramped high again to inhibit hopping and freeze all occupations. Finally, we employ the bandmapping technique to obtain the number of atoms on the even and odd sites~\cite{Trotzky12}. All bandmapping images are taken with $8 \,$ms time of flight and are imaged along the $y$-axis (i.e.\ orthogonal to the superlattice axis). 

The cloud size in the lattice is measured via in-situ imaging. It can be described by a gaussian, with standard deviation $\sigma$ of $\sim 14\,\mu$m and $\sim 11\,\mu$m in the $x$ and $y$ directions respectively. To estimate the number of sites in the central lattice tube, we take a value which corresponds to $4\, \sigma$, i.e.\ the range from $-2\, \sigma$ to $+2\, \sigma$.\\

\textit{Lattice Parameters.---}All lattice potentials result from retroreflected laser beams with gaussian intensity profiles and $1/e^2$ radii (waists) of $150\, \mu$m that are centered at the position of the atoms. The lattice beams along the $y$ and $z$ direction and the disorder lattice light originate from the same Coherent MBR-110 Ti:Sa laser operated at $738\,$nm, which is locked to its internal reference cavity. In order to eliminate cross-interference between beams along different axes, separate beams have orthogonal polarizations and are shifted to different frequencies using acousto-optic modulators. The relative phase $\phi$ of the disorder potential is controlled by slightly changing the frequency of this laser. The $1064\,$nm laser used for the long wavelength superlattice along the $x$-direction is locked to a dedicated reference cavity that is acoustically isolated and thermally stabilized to $65\,$kHz linewidth over $100\,$ms. A $532\,$nm Coherant Verdi laser is used as the short lattice and is locked to frequency doubled light from the $1064\,$nm laser via an offset lock.\\

\textit{Disorder Strength.---}The disorder strength $\Delta$ depends on the lattice depth of the $532\,$nm laser (main lattice), the lattice depth of the disorder lattice and the ratio of their wavelengths $\beta \approx 532/738$. We use $8\,E_{r,532}$ as the main lattice depth, where $E_{r,532} = h \times 17.64\,$kHz is the recoil energy in the $532\,$nm lattice. The disorder strength is then given by $\Delta/J = 6.67\,s_d$, where $s_d$ is the depth of the disorder lattice in units of the $738\,$nm recoil energy.\\

\begin{figure}
\centering
\includegraphics[width=84mm]{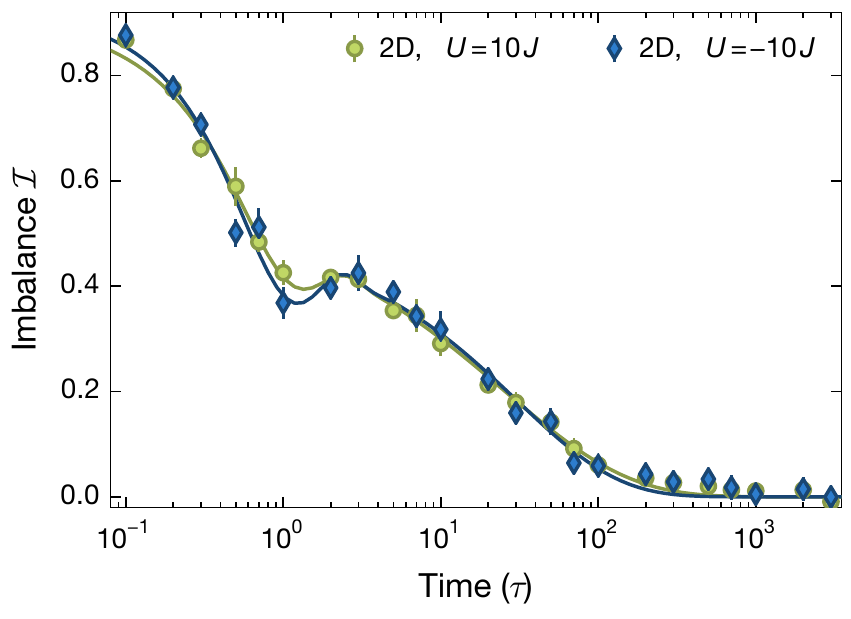}
\caption{\small{Dynamical Symmetry of the Hubbard Model: Time traces showing the evolution of the imbalance $\mathcal{I}$ in 2D for $\Delta=5\,J$ and $U=\pm 10\,J$. The data is identical within experimental scatter, showing that the decay of the imbalance does not depend on the sign of $U$. Each point is averaged over six disorder realizations $\phi$. Error bars denote the error of the mean. Solid lines are fitted stretched exponentials as per eq.~\ref{fit_func_Imb}.}}
\label{Hubbard_Symmetry_2D}
\end{figure}

\begin{figure}
\centering
\includegraphics[width=84mm]{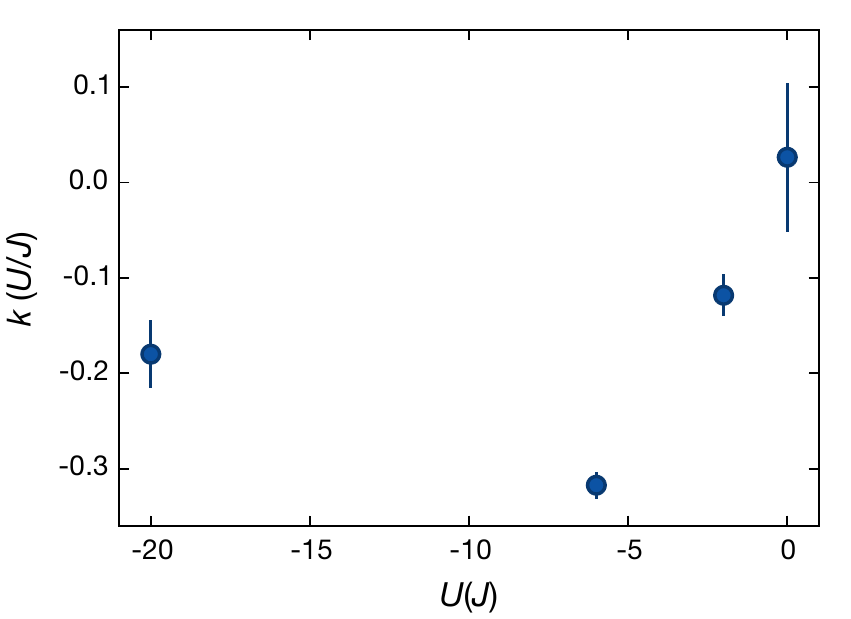}
\caption{\small{Fitted Power Law Exponents for 1D-2D Crossover: Shown are the power law exponents of the 1D-2D crossover as a function of interaction strength. Error bars denote the fit uncertainty.}}
\label{power_law_exponents}
\end{figure}

\textit{Dynamical $U \leftrightarrow -U$ Symmetry.---}We load the atoms in the lattice at magnetic fields above the Feshbach resonance at $202.1 \,G$~\cite{Regal03}. On this side of the resonance, the maximum positive scattering length that can be obtained with our experiment is $a\approx142 \, a_0$. This scattering length corresponds to an interaction strength of $U\approx 12 \, J$ in the 2D case. However, there is no such restriction in the case of negative interactions. Hence we measure most of the data with attractive interactions (i.e.\ $U$ negative). From a dynamical symmetry of the Hubbard model~\cite{Schneider12}, we expect the dynamics to be identical for $\pm U$, as was observed to be the case for the short time behavior in the 1D$^+$ case~\cite{Schreiber15}. However, to investigate whether the symmetry also holds for the long term dynamics, we measure the time traces in the 2D case for $U=\pm10\,J$. This data is shown in Fig.~\ref{Hubbard_Symmetry_2D} and indicates that the relaxation dynamics are indeed symmetric about $U=0$. Since fluctuations of the magnetic field would cause much smaller changes of the scattering length in the $U=+10 \, J$ case than in the $U=-10 \, J$ case, the observed $U\leftrightarrow-U$ symmetry indicates that neither field fluctuations nor the proximity to the Feshbach resonance contribute significantly to the decay of imbalance in the interacting 2D cases.\\

\textit{Doublon Dependence.---}While all the data presented in this paper is taken with a rather small doublon fraction of $\approx 8\, \%$, we also briefly investigate the effect of doublons on the imbalance lifetimes. We measure the lifetimes in the 1D$^+$ and the 2D case at $\Delta=5\,J,\,U=-6\,J$ for doublon fractions of $8\, \%$ and $50\, \%$ as well as for $U=-20\,J$ with $0\, \%$ and $8\, \%$ doublons. We can vary the doublon fraction between $\approx 8 \, \%$ to $\approx 50 \, \%$ by varying the scattering length during lattice loading between $a_{\text{load}} = -140\,\text{a}_{\text{0}}$ to $+140\,\text{a}_{\text{0}}$. In addition, we can reduce the doublon fraction further to almost zero by applying a near resonant blast pulse in the deep lattice. This removes the leftover doublons while leaving the singly occupied sites unaffected. For all of the above cases, we observe less than 5$\, \%$ change in the imbalance lifetime, with slightly longer imbalance lifetimes for higher doublon fractions. \\

\textit{Power-Law Fits for 1D-2D Crossover.---}The solid lines shown in Fig.~\ref{crossover_summary_fig} are fits to $T_{1/e} = b(J_\perp/J)^k$, where amplitude $b$ and the power law exponents $k$ are free fit parameters. The fitted exponents are plotted in Fig.\ \ref{power_law_exponents} as a function of interactions. Note that the exponents for all the interacting cases are negative with absolute values much smaller than unity. The exponents depend non-monotonously on the interaction strength, resembling the shape of the 2D curve in Fig.~\ref{const_times_fig}.\\

\begin{figure}
\centering
\includegraphics[width=84mm]{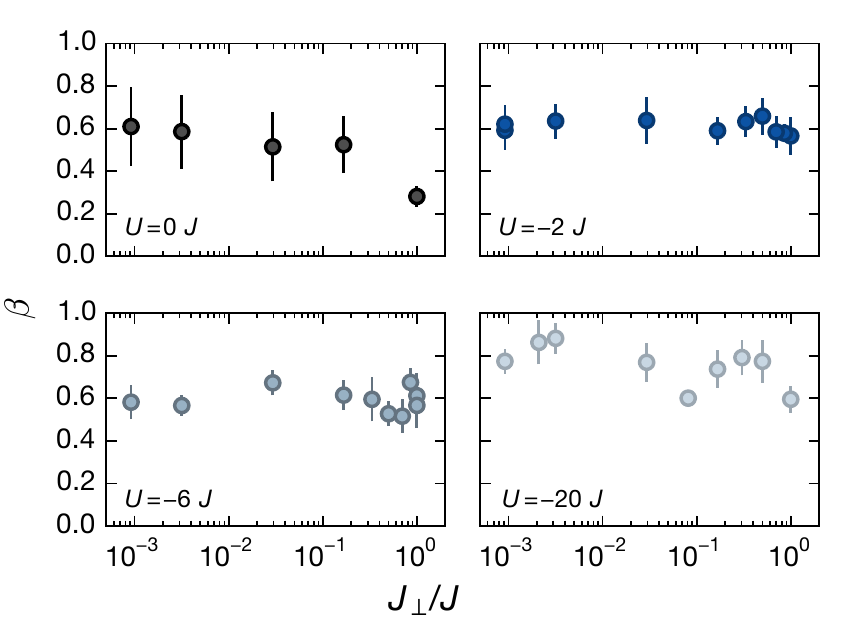}
\caption{\small{Variation of the stretching exponents $\beta$ in the 1D-2D Crossover: Shown are the fitted stretching exponents $\beta$ in the 1D-2D crossover as a function of interaction strength. Error bars denote the fit uncertainty.}}
\label{crossover_betas}
\end{figure}

\textit{Variation of the stretching exponents in the 1D-2D crossover.---} The fitted stretched exponents $\beta$ are shown in Fig.~\ref{crossover_betas} and typically lie in the range $[0.5,1]$. We do not observe a systematic dependence of the stretching exponent on the transverse coupling $J_\perp$. However, the exponents seem to be systematically larger in the case of $U=-20\,J$.\\

\end{document}